\documentclass[twocolumn,english]{IEEEtran}
\usepackage[T1]{fontenc}
\usepackage{babel}
\usepackage{textcomp}
\usepackage{amsmath}
\usepackage{graphicx}
\usepackage[unicode=true,
 bookmarks=true,bookmarksnumbered=true,bookmarksopen=true,bookmarksopenlevel=1,
 breaklinks=false,pdfborder={0 0 0},backref=false,colorlinks=false]
 {hyperref}
\hypersetup{pdftitle={Your Title},
 pdfauthor={Your Name},
 pdfpagelayout=OneColumn, pdfnewwindow=true, pdfstartview=XYZ, plainpages=false}

\makeatletter
 \let\oldforeign@language\foreign@language
 \DeclareRobustCommand{\foreign@language}[1]{%
   \lowercase{\oldforeign@language{#1}}}

\ifCLASSOPTIONcompsoc
\usepackage[caption=false,font=normalsize,labelfont=sf,textfont=sf]{subfig}
\else
\usepackage[caption=false,font=footnotesize]{subfig}
\fi

\usepackage{lineno}

\makeatother

\begin{document}

\title{Passive Mode-Locking and Tilted Waves in Broad-Area Vertical-Cavity
Surface-Emitting Lasers }

\author{M.~Marconi, J. Javaloyes,~\IEEEmembership{Member,~IEEE}, S. Balle,~\IEEEmembership{Member,~IEEE,}
and~M. Giudici~\IEEEmembership{Member,~IEEE}.%
\thanks{M. Marconi and M. Giudici are with the Institut Non Linéaire de Nice,
Université de Nice Sophia Antipolis - Centre National de la Recherche
Scientifique, 1361 route des lucioles, F-06560 Valbonne, France, e-mails:
\protect\href{http://mathias.marconi@inln.cnrs.fr}{mathias.marconi@inln.cnrs.fr}
and \protect\href{http://massimo.giudici@inln.cnrs.fr}{massimo.giudici@inln.cnrs.fr}.%
}%
\thanks{J. Javaloyes is with the Departament de Fisica, Universitat de les
Illes Baleares, C/ Valldemossa, km 7.5, E-07122 Palma de Mallorca,
Spain, e-mail: \protect\href{http://julien.javaloyes@uib.es}{julien.javaloyes@uib.es}%
}%
\thanks{S. Balle is with the Institut Mediterrani d'Estudis Avançats, CSIC-UIB,
E-07071 Palma de Mallorca, Spain, e-mail: \protect\href{http://salvador@imedea.uib-csic.es}{salvador@imedea.uib-csic.es}%
}}

\markboth{Journal of Selected Topics in Quantum Electronics}{Marconi \MakeLowercase{\emph{et al.}}:
Your Title}
\maketitle
\begin{abstract}
We show experimentally and theoretically that an electrically biased
$200\,\mu$m multi-transverse mode Vertical-Cavity Surface-Emitting
Laser can be passively mode-locked using optical feedback from a distant
Resonant Saturable Absorber Mirror. This is achieved when one cavity
is placed at the Fourier plane of the other. Such non conventional
optical feedback leads to the formation of two tilted plane waves
traveling in the external cavity with opposite transverse components
and alternating in time at every round-trip. Each of these plane waves
gives birth to a train of mode-locked pulses separated by twice the
external cavity round-trip, while the two trains are time shifted
by a round-trip. A large portion of the transverse section of the
device contributes to mode-locked emission leading to pulses of approximately
1~W peak power and $10\,$ps width. We discuss how inhomogeneities
in the transverse section of the saturable absorber select the emitted
tilted waves, thus leading to tunable emission over 4~nm.\end{abstract}

\section{Introduction}

\IEEEPARstart{L}{aser} mode-locking (ML) is a fascinating self-organized
cooperative effect involving a large number of longitudinal modes
that was recently linked to out-of-equilibrium phase transitions \cite{GP-PRL-02}.
From a practical point of view many applications require sources of
short pulses like e.g. medicine, metrology and communications \cite{haus00rev}.
Passive ML (PML) is arguably one of the most elegant method to obtain
such pulses. It is achieved by combining two elements, a laser amplifier
providing gain and a saturable absorber acting as a pulse shortening
element. Under appropriate conditions, the different dynamical properties
of the absorption and of the gain favor pulsed emission by creating
a limited time window for amplification around an intensity pulse
\cite{haus75f,haus75s}. PML can also be achieved using artificial
absorbers like e.g. nonlinear polarization rotation \cite{DLY-JQE-03},
Kerr lens mode-locking \cite{ippen94}, Crossed-Polarization \cite{JMB-PRL-06}
or Stark effect modulation \cite{WMD-OL-08}. The PML mechanism has
led to the shortest and most intense optical pulses ever generated
and pulses in the femto-second range are produced by dye \cite{FSY-JQE-83}
and solid-state lasers \cite{keller96}. However, the large size of
these devices and the need for optical pumping strongly limit their
application domain. 

More compact solutions can be envisaged using semiconductor devices:
PML is obtained in monolithic edge-emitting semiconductor lasers which
have the advantage of being electrically biased and to operate at
high repetition rates ($1\sim160$ GHz) although the peak powers that
can be obtained are usually limited because of Catastrophic Optical
Damage (COD) \cite{avrutin00}. Large output peak power in the kilowatt
range is commonly achieved by coupling Vertical-External-Cavity Surface-Emitting
Lasers (VECSEL) with a Semiconductor Saturable Absorber Mirror \cite{RWM-OE-10,WTB-OE-13}.
The external cavity is designed to operate in the fundamental Gaussian
mode while a large section of the VECSEL is optically pumped to achieve
large power, in this configuration the external cavity length leads
to repetition rates from a few to tens of GHz. In both monolithic
and external-cavity schemes, the presence of higher order transverse
modes of the resonator is usually perceived as detrimental for mode-locking
stability and it is avoided by cavity design. In fact, when several
higher order modes are present, the emission profile is usually not
stationary \cite{Fischer_1996}, and even chaotic filamentation may
occur \cite{Thompson_1972}. This is due to thermal effects imposing
a current-dependent refractive index profile, and to the so-called
Spatial Hole Burning (SHB). This phenomenon occurs in regions of high
optical intensity, where the local gain (and thus the local carrier
density) is depressed by stimulated emission, hence leading to a local
increase of refractive index which contributes to strengthen light
confinement and to further increase the local field intensity. On
the other hand, the possibility of achieving a cooperative effect
of transverses modes where they would contribute coherently to longitudinal
mode locking is very attractive for increasing the pulse power, since
it would allow to circumvent COD. 

In this work we propose a scheme for achieving mode-locking using
an electrically biased, $200\,\mu$m section Vertical-Cavity Surface-Emitting
Laser (VCSEL). This device is mounted in an external cavity configuration
closed by a Resonant Saturable Absorber Mirror (RSAM). Mode-locking
is obtained when placing the RSAM in the exact Fourier transform plane
of the VCSEL near-field profile, i.e. when imaging the VCSEL far-field
profile onto the RSAM. As a consequence, the VCSEL profile is imaged
onto itself after a single external cavity round-trip, but inverted.
This corresponds to a transverse magnification of -1. We show that
such configuration leads to the generation of two tilted plane waves
traveling in the external cavity with an opposite transverse component
and alternating each other at every round-trip. Each one of these
plane waves gives birth to a train of mode-locked pulses separated
by twice the external cavity round trip ($2\tau_{e}$), while the
two trains are time shifted of $\tau_{e}$. Almost the entire transverse
section of the VCSEL contributes to mode-locking leading to pulses
of 10~ps width and peak power around $1\,$W. We analyze the mechanism
leading to the selection of the tilted waves and we demonstrate a
technique for tuning the central wavelength of the mode-locked emission.
Our experimental findings are confirmed by a spatially resolved model
for the VCSEL and the RSAM taking into account the multiple reflections
in the external cavity and our specific imaging conditions.

\section{Experiment}

The VCSEL is a $980\,$nm device manufactured by ULM Photonics \cite{701502}.
Its standalone threshold current ($J_{st}$) is about $380\,$mA,
though emission is localized only at the external perimeter of the
device up to $J=850\,$mA, after which roll-off starts to occur. The
980 nm RSAM (BaTop Gmbh) has a transverse dimension of $4\times4$
mm$^{2}$ and it exhibits a low unsaturated reflectivity of 1\% that
increases up to $60\,\%$ when saturated. The RSAM saturation fluence
is $15\,\mu$J.cm$^{-2}$. These values are obtained at the RSAM resonant
wavelength which can be thermally tuned over $3\,$nm (between $T_{1}=10\,\text{\textdegree C}$
and $T_{2}=50\,\text{\textdegree C}$). The Full Width at Half Maximum
(FWHM) of the RSAM resonance is around $16\,$nm and the saturable
absorption recovery time is around $1\,$ps.

\begin{figure}[h]
\centering{}\includegraphics[bb=0bp 0bp 396bp 248bp,clip,width=0.48\textwidth]{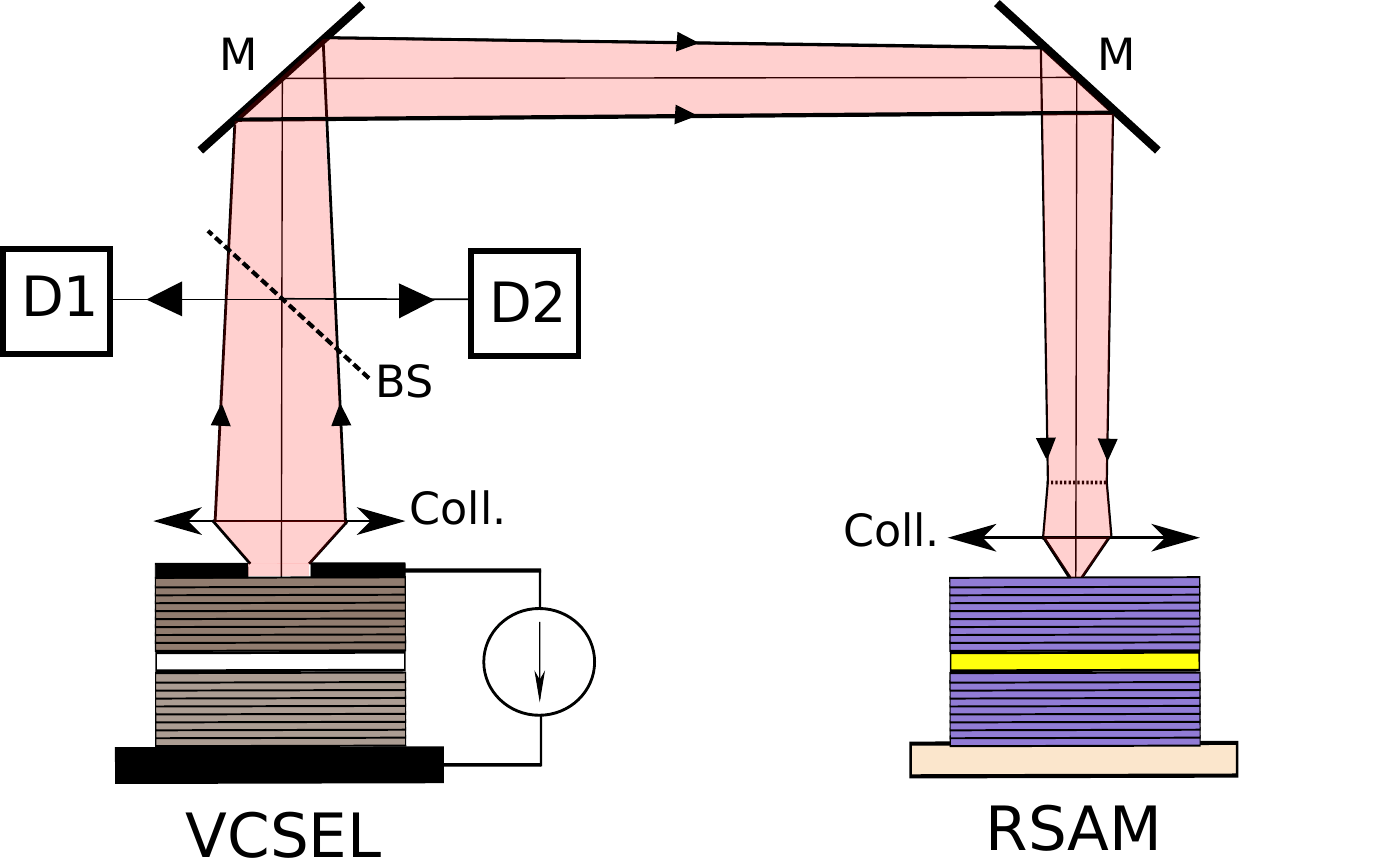}
\caption{Experimental Set-up: Temperature-stabilized VCSEL and RSAM. Coll.:
Aspheric Lens, BS : Beam Splitter, M: Mirror and D1/D2: Detectors
and CCD cameras.\label{setup}}
\end{figure}

\begin{figure}[t]
\begin{centering}
\includegraphics[width=0.5\textwidth]{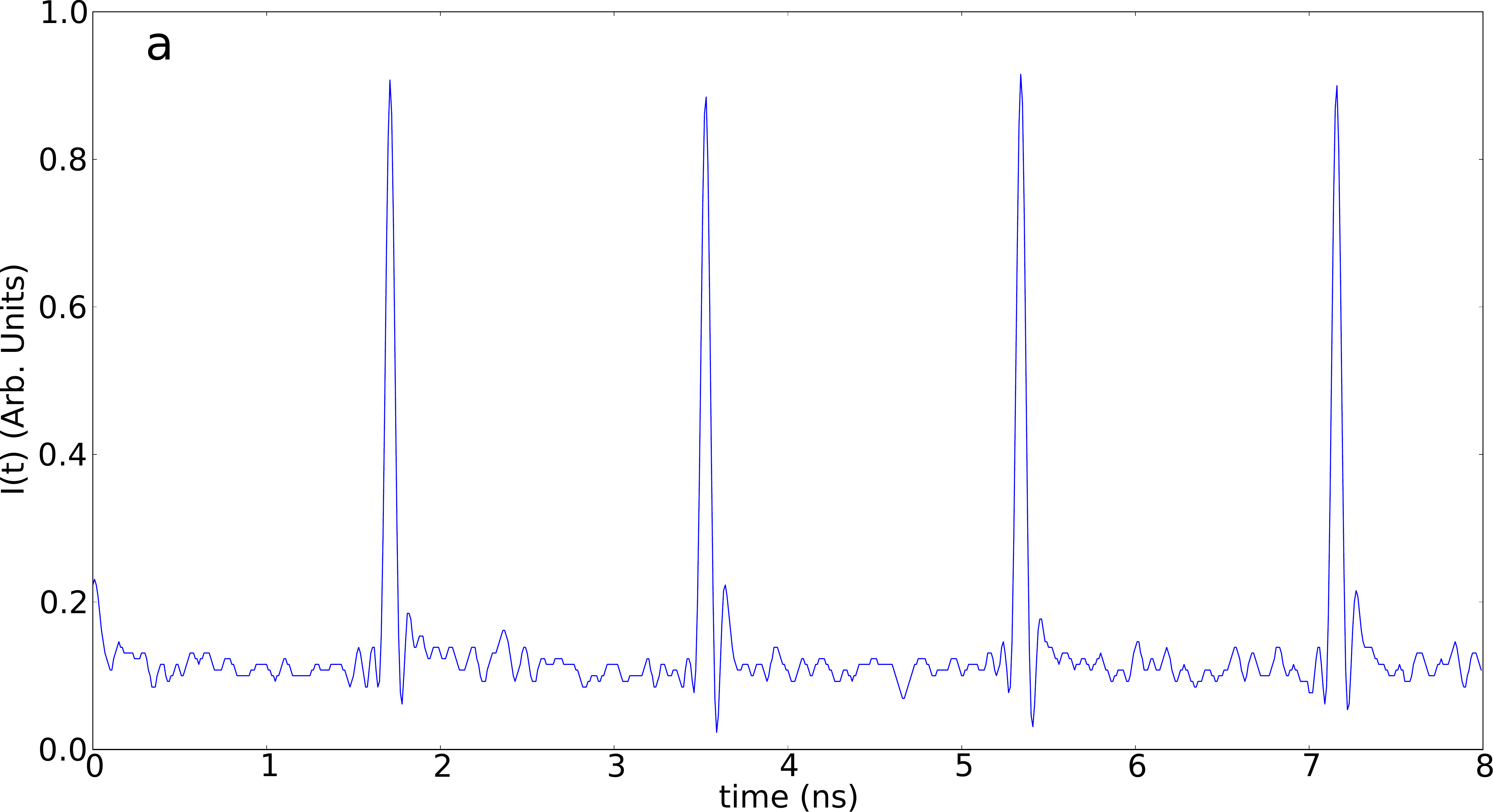}
\par\end{centering}

\vspace{0.1cm}

\centering{}\includegraphics[width=0.5\textwidth]{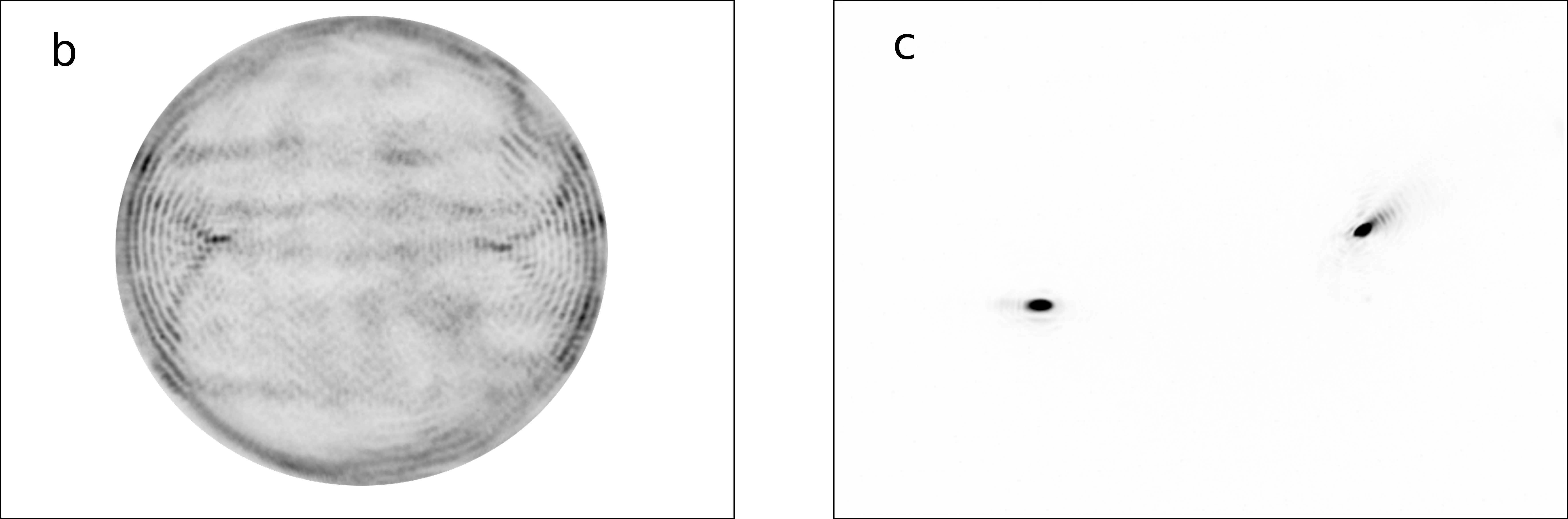}
\caption{Panel a): Temporal trace of the VCSEL in the mode-locked regime. Panel
b): Corresponding Near-Field emission of VCSEL. Panel c): Corresponding
Far-field emission from the VCSEL. Intensity grows from white to black.
$J$=600~mA.\label{modelocking} }
\end{figure}

The set up is shown in Fig.~\ref{setup}. Both the VCSEL and RSAM
are mounted on temperature controlled substrates which allow for tuning
the resonance frequency of each cavity; parameters are set for having
the emission of the VCSEL resonant with the RSAM. The light emitted
by the VCSEL is collected by a large numerical aperture (0.68) aspheric
lens and a similar lens is placed in front of the RSAM. A $10\,\%$
reflection beam splitter allows for light extraction from the external
cavity and to monitor both the VCSEL and the RSAM outputs. Intensity
output is monitored by a $33\,$GHz oscilloscope coupled with fast
$10\,$GHz detector. Part of the light is sent to two CCD cameras;
the first one records the near-field profile of the VCSEL, while the
second records the VCSEL's far-field profile. The light reflected
by the RSAM is also used for monitoring and a third CCD camera records
the light on the RSAM surface. The external cavity length is fixed
to $L=30\,$cm. One of the most important parameters for achieving
mode-locking in this setup is the imaging condition of the VCSEL onto
the RSAM. We obtain mode-locking when the RSAM is placed in the plane
where the exact Fourier transform of the VCSEL near-field occurs.
This working condition is obtained by imaging the VCSEL near-field
profile onto the front focal plane of the aspheric lens placed in
front of the RSAM, while the RSAM is placed onto the back focal plane
of this lens. We remark that this leads to a non-local feedback from
the RSAM onto the VCSEL: if the RSAM were a normal mirror, the VCSEL
near-field profile would be inversely imaged onto itself after a cavity
round-trip.

Panel a) in Fig.~\ref{modelocking} displays the time trace of the
VCSEL in the mode-locking regime which consists of a regular train
of pulses with a period equal to the round-trip time in the external
cavity $\tau_{e}=2L/c=2\,$ns. The pulse width cannot be determined
from the oscilloscope traces, which are limited by our real-time detection
system ($10\,$GHz effective bandwidth). However, an estimate of the
pulse width can be obtained from the optical spectrum of the output,
which exhibits a broad spectral peak whose FWHM is around $0.12\,$nm
that corresponds, assuming a time-bandwidth product of $0.4$, to
a pulse width of $10\,$ps FWHM. The pulse was also detected by a
42 GHz detector, which confirms a pulse width of less than $12\,$ps
FWHM considering the oscilloscope bandwidth limit.

Panels b) and c) in Fig.~\ref{modelocking} show the time-averaged
near-field and far-field profiles of the VCSEL, respectively. In addition,
we verified that the image of the RSAM surface (not shown) is very
similar to Fig.~\ref{modelocking}c), thus revealing that the RSAM
is effectively placed in the Fourier plane of the VCSEL's near-field.
The far-field of the VCSEL exhibits two bright, off-axis spots that
indicate the presence of two counter-propagating tilted waves along
the cross section of the VCSEL. The transverse wave vector of each
of these waves is related in direction and modulus to the position
of the spots, and their symmetry with respect to the optical axis
indicates that the two transverse wave vectors are one opposite to
the other.

\begin{figure}[t]
\centering{}\includegraphics[clip,width=8.5cm]{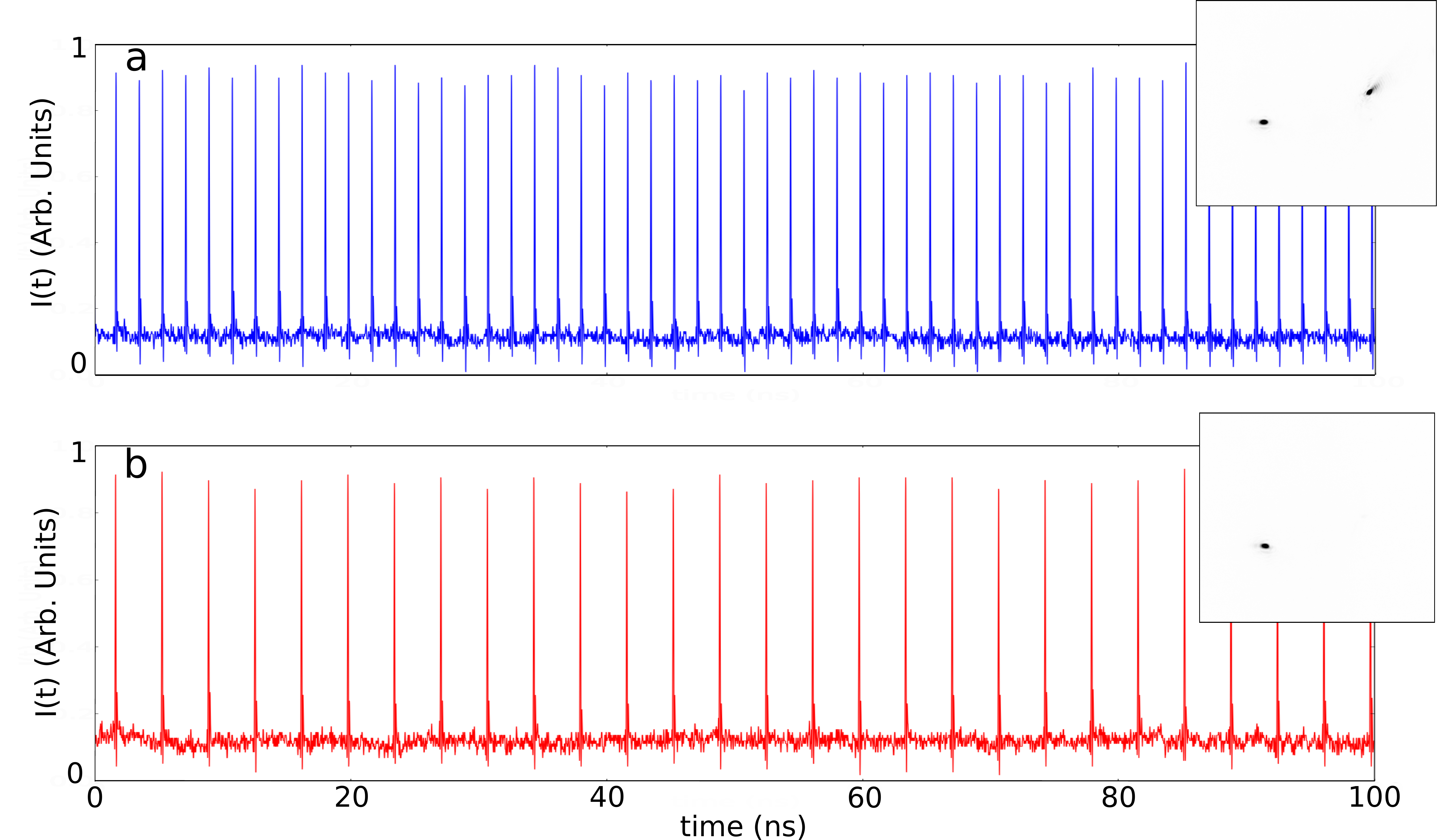} \caption{Temporal signal obtained when detecting simultaneously the two spots
(panel a) of the far-field emission of the VCSEL or a single spot
(panel b). The insets represent the detected spots in the far-field.
\label{twospots} }
\end{figure}

Remarkably, however, no interference pattern is visible in the near-field
emission of the VCSEL. The concentric rings close to the limit of
the VCSEL arise from current crowding \cite{BLA-PRE-04} and do not
contribute significantly to the bright spots in the far-field, as
it was verified by filtering them out. The lack of interference pattern
between the counter-propagating waves in the time-averaged near field
implies that these waves are not simultaneously present; they rather
should alternate each other in time. This is apparent in Fig.~\ref{twospots},
where the time trace obtained from the whole far-field is compared
to the one obtained by detecting only one of the two bright spots.
While the former trace has the characteristics discussed above, the
latter trace consists of a periodic train of pulses at a period $2\tau_{e}$.
Thus the trace from the whole far-field is obtained by interleaving
two identical pulse trains of period $2\tau_{e}$ with a time delay
$\tau_{e}$ one with respect to the other, each train corresponding
to a tilted wave with opposite transverse wave vector. 

\begin{figure}[t]
\centering{}\includegraphics[clip,width=8.5cm]{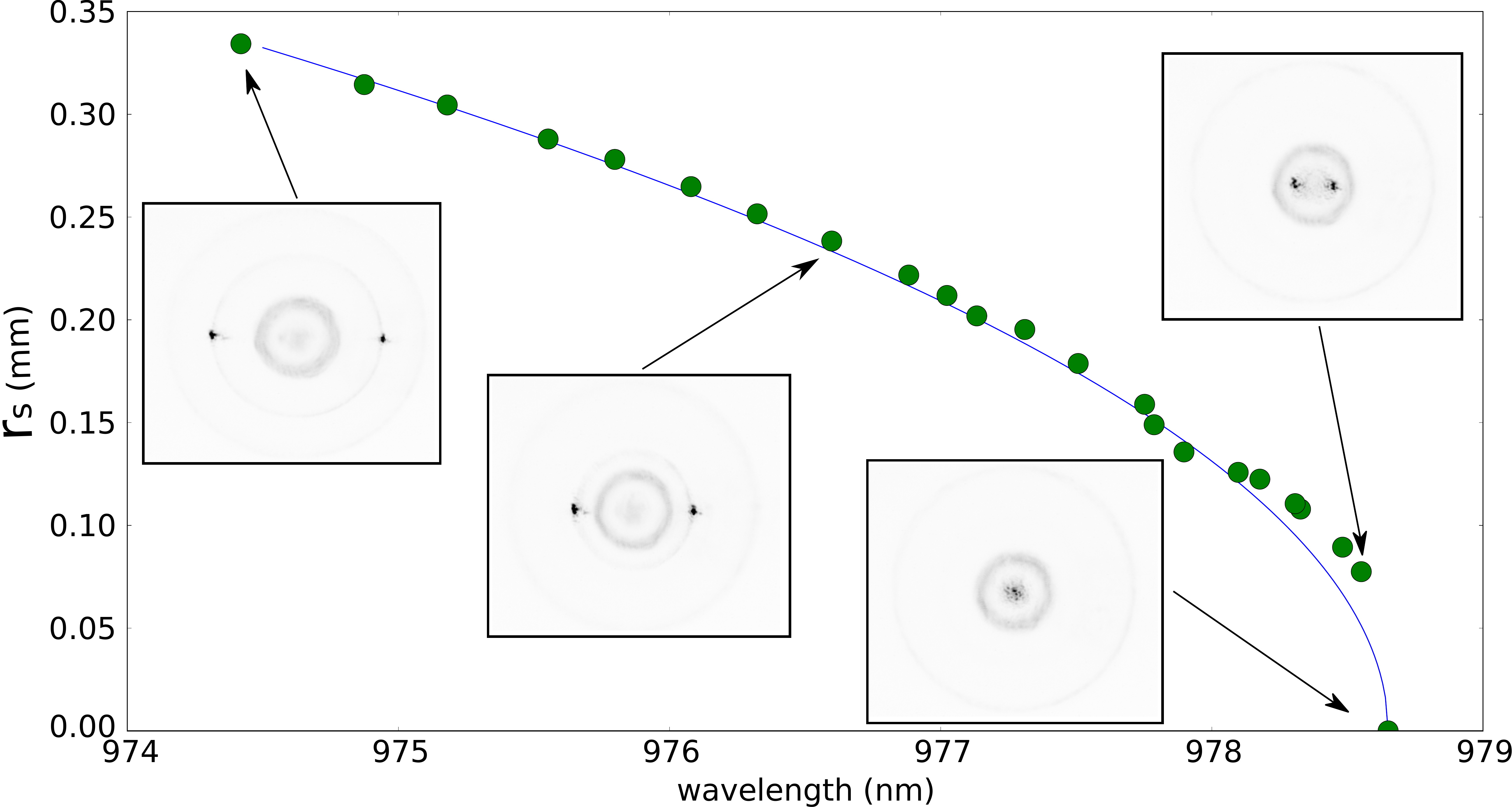} \caption{Off-axis position of a single spot in the far-field profile as a function
of the spectral emission peak of the VCSEL. Far-field profile is shown
for a discrete number of points in the graph. VCSEL emission is in
the mode-locking regime for all green points of the graph while the
one at the highest wavelength, where a single on-axis spot appears
in the far-field profile, corresponds to an irregular dynamics. VCSEL
is biased at 700~mA. The transverse wave vector is selected by laterally
shifting the RSAM along the back focal plane of its collimating lens.
The large size of the RSAM section ( $4\times4\,$mm) with respect
to the far-field size (0.7~mm at most) renders this operation feasible.
The blue line has been obtained plotting $r_{s}=\frac{\lambda_{0}f}{2\pi}\sqrt{\lambda^{-2}-\lambda_{0}^{-2}}$,
with $\lambda_{0}=978.65\,$nm and $f=8\,$mm. A similar tuning curve
is obtained by laterally shifting the collimating lens of the RSAM.
\label{tilted_ML} }
\end{figure}

This mode-locked dynamics does not depend critically on the transverse
wave vector value selected by the system. Such value can be modified
by shifting the RSAM device laterally (i.e. along the back focal plane
of its collimating lens) or slightly displacing the collimating lens
off the axis defined by the centers of the VCSEL and the RSAM. The
change in selected transverse wave vector is evidenced by the variations
in the separation between the two spots on the RSAM mirror, as shown
in Fig.~\ref{tilted_ML}. The position of each spot on the RSAM with
respect to the optical axis is related to the transverse wave vector
$\vec{K}_{\bot}=\vec{K}-\vec{K}_{0}$ of the plane wave by 
\[
\vec{r}_{s}=\lambda f\frac{\vec{K}_{\bot}}{2\pi}\;,
\]
where $f$ is the focal length of the lens in front of the RSAM, $\lambda$
is the wavelength of the light, $\vec{K}$ is the wave vector emitted
by the VCSEL while $\vec{K}_{0}$ is its component along the cavity
axis. Although substantial changes in emission wavelength ---over
4 nm tuning--- can be induced in this way, it is observed that within
a wide parameter range, the temporal characteristics of the pulse
train do not change. This fact opens very interesting possibilities
in terms of wavelength tuning and of beam stirring of a mode-locked
emission. The only noticeable effect of varying the position of the
spot is a slight reduction of the pulse peak power. We attribute this
effect to the increased losses experienced in the external cavity
by wave vectors with large $K_{\bot}$ as a result of the DBRs reflectivity
angular dependence and/or the finite numerical aperture of the collimating
lenses. This point will be further discussed in the theoretical section.

Beyond the maximal separation of the two spots on the RSAM shown in
Fig.~\ref{tilted_ML}, mode-locking is suddenly lost. Importantly,
mode-locking strongly deteriorates in regularity when the two spots
are brought to coincide, leading, in some cases, to CW emission. Hence,
in our setup, regular mode-locking was not achieved with $K_{\perp}\simeq0$,
i.e. for a plane-wave emission almost parallel to the optical axis
of the VCSEL. 

\begin{figure}[t]
\centering{}\includegraphics[clip,width=8.5cm]{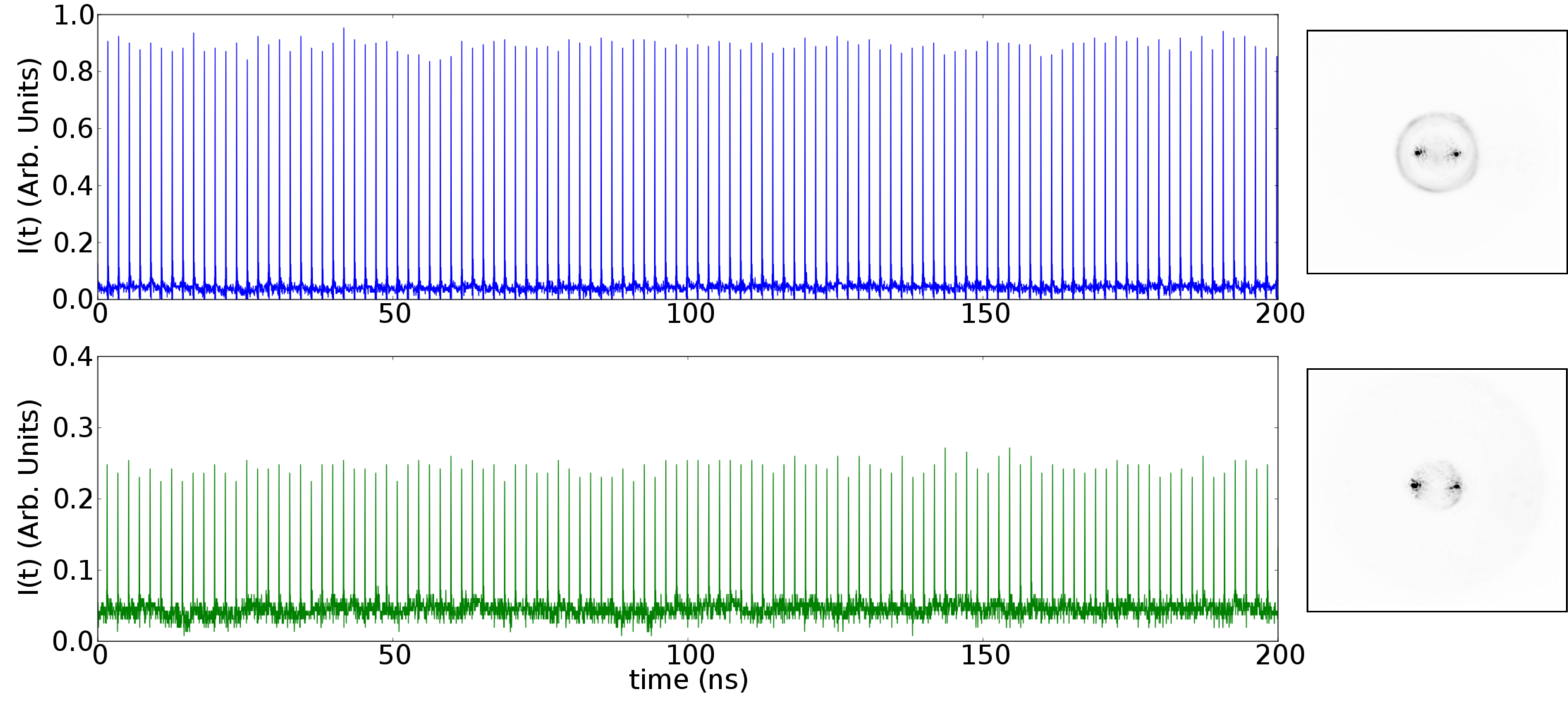} \caption{VCSEL time traces emission and corresponding far-field emission profile
in the mode-locking regime for two values of the VCSEL bias. The upper
and the lower panels correspond to $J=703\,$mA and $J=285\,$mA.
\label{vsJ} }
\end{figure}

The mode-locking regime is stable in a very broad range of the VCSEL
current, namely $285\,$mA$<J<703\,$mA. If the bias current is varied
within this range while keeping the alignment, the separation of the
two bright spots in the far-field profile remains constant, see Fig.~\ref{vsJ}.
However the spectral peak corresponding to the mode-locking emission
redshifts from 976~nm ($J=285\,$mA) up to 978~nm ($J=703\,$mA)
due to Joule heating of the VCSEL. Therefore, the selection of $K_{\bot}$
does not depend on the detuning between the two cavities, at least
in the range spanned. 

We also found that, for the external cavity length considered, the
PML regime is bistable with the off solution for $J<J_{st}$, see
\cite{MJB-PRL-14} for details. In these conditions and after setting
the system in the off solution, we were able to start PML emission
by perturbing optically the RSAM section at the point where one of
the two spots appears when the system operates in the PML regime.
The local perturbation has been realized by injecting an external
coherent beam tuned with the RSAM cavity resonance and having a waist
diameter of less than $10\,\mu$m.

\section{Discussion of the experimental evidences}

These observations disclose a possible explanation about the emergence
of the mode-locking regime in our system. From the dynamical point
of view, mode-locking is favored when the saturable absorber is more
easily saturated than the amplifier \cite{haus00rev}. In our scheme,
this is achieved when the VCSEL emits a tilted plane wave, which imposes
a low power density (hence low saturation) in the gain section and,
at the same time, a strong local saturation of the RSAM. The reason
is that the VCSEL and the RSAM lie one into the Fourier plane of the
other and in these conditions any plane wave emitted by the VCSEL
yields a spot on the RSAM and vice-versa. On the other hand, the unsaturated
reflectivity of the RSAM is very low ($\sim1\,\%$) at resonance,
rising up to $\sim60\,\%$ when fully saturated. Hence, if the VCSEL
emits a pulse in the form of a plane wave with a transverse wave vector
$\vec{K}_{\bot}$, all the power will concentrate on a single spot
at $\vec{r}_{s}$ on the RSAM, hereby strongly saturating the RSAM
which becomes reflective. The light thus comes back to the VCSEL,
where it arrives with a transverse component $-\vec{K}_{\bot}$. Upon
amplification and reflection at the VCSEL, the process repeats for
this plane wave with opposite transverse wave vector, which is now
imaged onto the RSAM at $-\vec{r}_{s}$. Thus, after two round-trips
the original wave at $\vec{K}_{\bot}$ overlaps with itself, leading
to a pulse train at twice the round-trip time for this wave which
is alternating in time with a delay of one round-trip with its replica
at $-\vec{K}_{\bot}$. 

This explains the observed dynamics, and also the lack of interference
pattern in the form of rolls in the VCSEL near-field: even if two
tilted waves are propagating within the VCSEL resonator, they alternate
in time with a delay of one round-trip in the external cavity, so
that they never coexist and do not yield the expected interference
pattern. The two spots appearing simultaneously in the far-field (i.e.
over the RSAM) are only an artifact of the time-averaging operated
by the CCD camera. At variance with tilted waves leading to narrow
spots onto the RSAM, the flower mode present on the external frontier
of the VCSEL does not contribute to PML. In fact, the Fourier Transform
of such flower mode is a radial Bessel function of large order $\sim J_{m}\left(r\right)$
with $m\gg1$, which corresponds to an extended ring profile. In this
case, the power density on the RSAM remains always below the saturation
fluence of the RSAM. 

While the above scenario is valid for a large interval of values of
the transverse wave vector $\vec{K}_{\bot}$, experimental results
show that the system selects a well-defined value when operating in
the PML regime. In principle, this selection may arise from several
factors. On the one hand, both the VCSEL and the RSAM are Fabry-Perot
cavities defined by Bragg mirrors which complex reflectivity have
an angular dependence. Moreover, the reflectivity of a Fabry-Perot
cavity depends not only on the wavelength but also on the angle of
incidence of the light, and the resonant wavelength of the cavity
blue-shifts as the incidence angle increases. This effect can be further
enhanced in our system due to the compound-cavity effect. Nevertheless,
experimental evidences shown in Fig.~\ref{vsJ}, where the VCSEL
cavity resonance is varied over 2~nm while leaving the transverse
wave vector unchanged, seems to indicate that cavities detuning do
not play an important role in selecting the value of $\vec{K}_{\bot}.$
This can be understood when considering that the FWHM bandwidth of
the RSAM absorption curve (16~nm) is much larger than the VCSEL one
(1~nm).

On the other hand, wave vector selectivity may also arise from any
imperfections on the RSAM/VCSEL mirrors that break the spatial invariance.
State-of-the-art fabrication process does not fully prevent from formation
of small size (a few $\mu$m) defects in the transverse plane of semiconductor
micro-resonators. These are in general local spatial variations of
the semiconductor resonator characteristics \cite{OKS-OQEL-92}, which
in the case of RSAM would be mainly related to a local variation of
unsaturated reflectivity. Visual inspection of the reflection profile
of a transversally homogeneous injected monochromatic field onto several
used RSAMs has indeed shown the existence of these local variations.\textbf{
} Such inhomogeneities on the RSAM surface may correspond to locations
where the RSAM exhibits a lower unsaturated reflectivity, hence leading
the system to select the tilted wave which image on the RSAM surface
coincide with one of these defect. By laterally shifting the RSAM
along the back focal plane of its collimating lens, the position of
the defect changes in the far-field plane, thus selecting a tilted
wave with another transverse wave vector component. This leads to
a continuous scan of the distance between the two spots in the far-field,
as shown in Fig.~\ref{tilted_ML} and a corresponding change of the
emission wavelength, as expected for tilted waves. The experimental
evidence of Fig.~\ref{tilted_ML} seems to indicate that, in our
system, the presence of these inhomogeneities is ruling the selection
of the transverse wave vector. The role of small imperfections in
the RSAM structure in the transverse wave vector selection is confirmed
by the theoretical results in the next section.

\section{Theoretical results}

Our model for the dynamical evolution of the fields $E_{j}$ and the
normalized carrier density $N_{j}$ in the quantum well (QW) regions
of the VCSEL $(j=1)$ and the RSAM $(j=2)$ is deduced in detail in
the appendix, and it reads 
\begin{eqnarray}
\dot{E}_{1} & \negthickspace\negthickspace=\negthickspace\negthickspace & \left[\left(1-i\alpha_{1}\right)N_{1}-1+i\Delta_{\perp}+c_{1}\Delta_{\perp}^{2}\right]E_{1}+h_{1}Y_{1},\label{eq:E1}\\
\dot{E}_{2} & \negthickspace\negthickspace=\negthickspace\negthickspace & \left[\left(1-i\alpha_{2}\right)N_{2}-z+ib\Delta_{\perp}+c_{2}\Delta_{\perp}^{2}\right]E_{2}+h_{2}Y_{2},\label{eq:E2}\\
\dot{N}_{1} & \negthickspace\negthickspace=\negthickspace\negthickspace & \gamma_{1}\left[J_{1}-\left(1+\vert E_{1}\vert^{2}\right)N_{1}\right]+\mathcal{D}_{1}\Delta_{\perp}N_{1},\label{eq:D1}\\
\dot{N}_{2} & \negthickspace\negthickspace=\negthickspace\negthickspace & \gamma_{2}\left[J_{2}-\left(1+s\vert E_{2}\vert^{2}\right)N_{2}\right]+\mathcal{D}_{2}\Delta_{\perp}N_{2},\label{eq:D2}
\end{eqnarray}
where $\Delta_{\perp}=\partial_{x}^{2}+\partial_{y}^{2}$ is the transverse
Laplacian and $Y_{j}$ denotes the field injected in device $j$.
In Eqs.~(\ref{eq:E1}-\ref{eq:D2}) time and space are normalized
to the photon lifetime $\kappa_{1}^{-1}$ and the diffraction length
in the VCSEL $L_{d}$, respectively. The complex parameter $z$ is
decomposed as $z=a-i\delta$ where $a$ represents the ratio of the
photon decay rates between the VCSEL and the RSAM cavities and $\delta$
is the scaled detuning between the two cavity resonances. The scaled
carrier recovery rates and the biases are denoted $\gamma_{j}$ and
$J_{j}$ respectively. We define the ratio of the saturation intensities
of the VCSEL and the RSAM as $s$. The angular dependence from the
reflectivity of the DBR mirrors as well as the one stemming from the
Fabry-Perot cavities are contained in the transverse Laplacian and
the parameters $c_{1,2}.$ The RSAM being a broadband fast absorber,
it is characterized by $J_{2}<0$, $\gamma_{2}\gg\gamma_{1}$, $a\gg1$
as well as $s\gg1$.

\subsubsection{Injected fields and device coupling}

In the case of a self-imaging configuration, the link between the
two devices is achieved by expressing the emitted fields $O_{j}$
as a combination of the reflection of the injection fields $Y_{j}$
and the self-emission $E_{j}$, which reads
\begin{equation}
O_{j}\left(r,t\right)=\eta_{j}E_{j}\left(r,t\right)-Y_{j}\left(r,t\right),\label{eq:output_field}
\end{equation}
where we defined $\eta_{j}=t_{1}^{j}/\left(1+r_{1}^{j}\right)$ with
$t_{1}^{j}$ and $r_{1}^{j}$ the transmission and reflection coefficients
in amplitude of the emitting DBR from the inside to the outside of
the cavity. Considering the propagation delay and the losses incurred
by the beam-splitter allowing for the light extraction, the link between
the two devices in ensured by the following two delayed Algebraic
Equations \cite{JB-OE-12}
\begin{equation}
Y_{1}\left(r,t\right)=t_{bs}O_{2}\left(r,t-\tau_{e}\right),Y_{2}\left(r,t\right)=t_{bs}O_{1}\left(r,t-\tau_{e}\right),
\end{equation}
with $t_{bs}$ a complex number whose modulus and phase model the
losses induced by the beam-splitter and the single trip feedback phase,
respectively. 

However, the image of the VCSEL through its collimator is placed exactly
at the object focal plane of the collimator in front of the RSAM.
In turn, the RSAM is on the image focal plane of its collimator. In
this configuration, and assuming that the collimators have a large
diameter, the field injected in one device is the (delayed) Fourier
transform in space of the field coming from the other, which is modelized
via Kirchhoff's formula for a lens with focal length $f$ as
\begin{eqnarray}
Y_{1}(r,t) & \negthickspace\negthickspace=\negthickspace\negthickspace & t_{bs}\int\, O_{2}(r^{'},t-\tau_{e})e^{-i\frac{\omega_{0}}{c}\frac{r\text{\ensuremath{\cdot}r'}}{f}}d^{2}r'\equiv\mathcal{F}_{\tau}(O_{2})\label{eq:Y1}\\
Y_{2}(r,t) & \negthickspace\negthickspace=\negthickspace\negthickspace & t_{bs}\int\, O_{1}(r^{'},t-\tau_{e})e^{-i\frac{\omega_{0}}{c}\frac{r\text{\ensuremath{\cdot}r'}}{f}}d^{2}r'\equiv\mathcal{F}_{\tau}(O_{1})\label{eq:Y2}
\end{eqnarray}

\noindent and where the beam-splitter losses may contain a normalization
constant due to the Kirchhoff's integral. It is worth remarking that
Eqs.~(\ref{eq:Y1}-\ref{eq:Y2}) describe the coupling of the two
devices to all orders of reflection in the external cavity. They define
a delayed map where the field injected into one device returns inverted.
This can be seen, for instance, using Eq.~(\ref{eq:output_field})
and saying that the RSAM has an effective reflectivity such that $O_{j}=r_{eff}Y_{j}$
and substituting in Eq.~(\ref{eq:Y1}), which yields 
\begin{equation}
Y_{1}(r_{\perp},t)=\mu E_{1}(-r_{\perp},t-2\tau)
\end{equation}
where we have used that $\mathcal{F}_{\tau}\circ\mathcal{F}_{\tau}\left[Y_{2}\right]=\alpha Y_{2}\left(-r_{\perp},t-2\tau\right)$
and $\mu$ = t$_{bs}^{2}r_{eff}$ describes the combined effect of
the beam splitter and of the RSAM reflection. Thus, after one round-trip,
$Y_{1}$ overlaps with a spatially reversed copy of itself, and it
requires a second round-trip to achieve proper overlap.

\subsubsection{Parameters}

The reflectivities of the DBRs in the VCSEL and the RSAM are taken
as $\left(r_{1}^{\left(1\right)},r_{2}^{\left(1\right)}\right)=\left(\sqrt{0.942},1\right)$
and $\left(r_{1}^{\left(2\right)},r_{2}^{\left(2\right)}\right)=\left(\sqrt{0.59},1\right)$.
We assume that the single trip in the VCSEL and the RSAM is $\tau=30\,$fs
corresponding to an effective length $L_{z}=2.6\,\mu$m with a index
of $n=3.5$. This gives us $\kappa_{1}=10^{12}\,$rad.s$^{-1}$ and
$\kappa_{2}=10^{13}\,$rad.s$^{-1}$, yielding a FWHM for the resonances
of $\Delta\lambda_{j}=\kappa_{j}\lambda_{0}^{2}/\left(\pi c\right)$
of $1\,$nm and $10\,$nm respectively. Incidentally, we find that
$a=10$ and $\left(h_{1},h_{2}\right)=\left(2,20\right)$. The diffraction
length is found to be $L_{d}=\sqrt{L_{z}\left[q_{0}\left(1-r_{1}r_{2}\right)\right]^{-1}}=2\,\mu$m.
We define two numerical domains which are twice the size of the VCSEL
and of the RSAM which have both a normalized length $L_{\perp}=200$.
The other parameters are $\alpha_{1}=2$, $\alpha_{2}=0.5$, $b=10^{-2}$,$\gamma_{1}=10^{-3}$,
$\gamma_{2}=0.1$, $s=10$, $t_{bs}=0.9$, $\mathcal{D}_{1}=\mathcal{D}_{2}=10^{-3}$
$c_{1}=10^{-2}$ and $c_{2}=10^{-6}.$

\subsubsection{Numerical considerations}

The simulation of PML lasers is a very demanding problem from the
computational point of view: while pulses may form on a relatively
short time scale of a few tens of round-trips, the pulse characteristics
only settle on a much longer time scale \cite{JB-JQE-10}. If anything,
the complex transverse dynamics present in our case shall slow down
the dynamics even further. Considering a delay of $\tau_{e}=2\,$ns
and a time step of $\delta t=10^{-2}$ implies that one must keep
four memory buffers for $E_{j}$ and $Y_{j}$ of size $\left(\tau_{e}\kappa_{1}/\delta t\right)N_{x}N_{y}\sim2\times10^{3}N_{x}N_{y}$
with $N_{x}$ and $N_{y}$ the number of mesh points in the two transverse
directions. This amounts to $32\,$Gigabytes with $N_{x}=N_{y}=512$.
Notice in addition that such values of $N_{x,y}$ are not particularly
large if one considers that only half of the mesh discretization in
the VCSEL is electrically pumped, the rest being used only for letting
the field decay to zero and prevent aliasing, a common problem with
spectral methods in broad-area lasers, see for instance \cite{PJB-JSTQE-13}
for a discussion. The time step of $\delta t=10^{-2}$ is neither
particularly small if one considers the stiffness incurred by the
RSAM response. For these reasons we restrict our analysis to a single
transverse dimension. We believe that the main spatial feature being
lost by such simplification is the description of the non mode-locked
two dimensional radial flower mode \cite{BLA-PRE-04} on the border
of the VCSEL, which has already been qualitatively discussed. To conclude,
for the sake of simplicity, we used a time delay of $\tau_{e}=200\,$ps
(instead of $1\,$ns) to avoid the temporal complexity of our setup
(leading to localized structures, as discussed in \cite{MJB-PRL-14}),
and to concentrate only on the spatial aspects of the problem. We
integrated numerically Eqs.~(\ref{eq:E1}-\ref{eq:D2}) using a semi-implicit
split-step method where the spatial operators are integrated using
the Fourier Transform. Such semi-implicit method is particularly appropriate
to the stiffness induced by the broad response of the RSAM. The simulation
time was $500$ single-trips in the external cavity, i.e. $100\,$ns.

\subsubsection{RSAM characteristics}

\begin{figure}[t]
\centering{}\includegraphics[clip,width=0.5\textwidth]{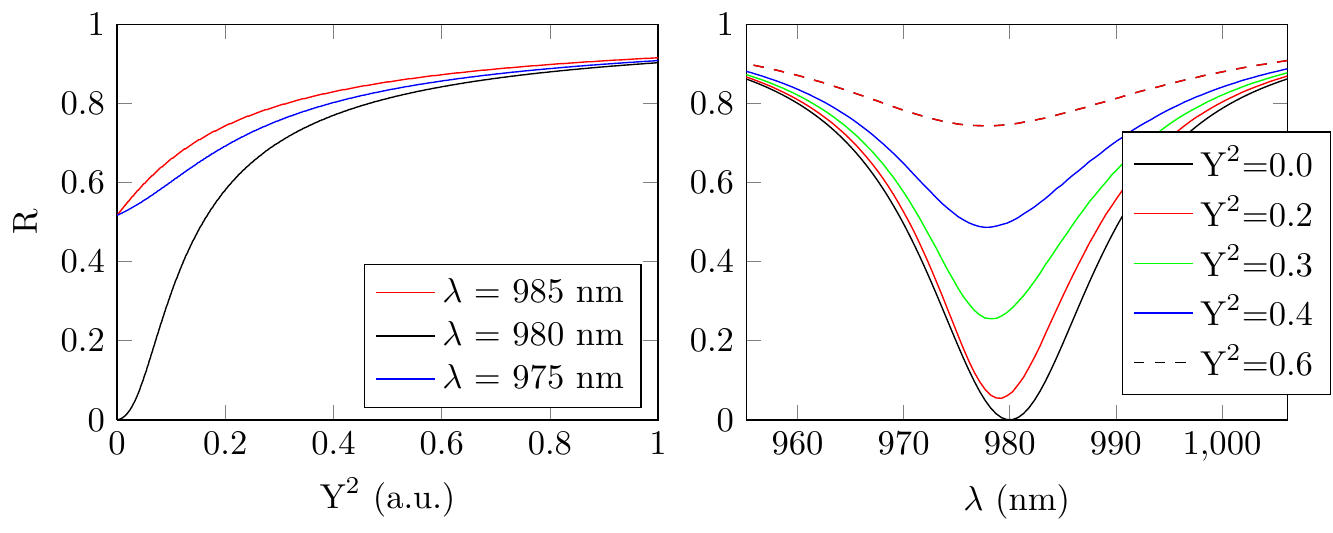}
\caption{Reflectivity as a function of the input field intensity (left) for
various values of the wavelength, the resonance being chosen to be
$\lambda_{0}=980\,$nm and (right) reflectivity as a function of wavelength
for increasing input power. \label{RSAM_theo} }
\end{figure}

The basic properties of the RSAM can be found assuming an incoming
plane wave of amplitude $Y$ at normal incidence and at frequency
$\omega$ i.e. $Y_{2}=Y\exp\left(-i\omega t\right)$. The response
of the RSAM is given by 
\begin{eqnarray}
\left(z-i\omega-\frac{J_{2}\left(1-i\alpha_{2}\right)}{1+s|E_{2}|^{2}}\right)E_{2} & = & h_{2}Y.\label{eq:RSAM}
\end{eqnarray}

The effective reflectivities are obtained setting $O_{2}=rY_{2}$
that is to say $r=E_{2}/Y_{2}-1$ yielding the unsaturated and saturated
responses as
\begin{equation}
r_{u}=\frac{h_{2}-z+i\omega+J_{2}\left(1-i\alpha_{2}\right)}{z-i\omega-J_{2}\left(1-i\alpha_{2}\right)},r_{s}=\frac{h_{2}-z+i\omega}{z-i\omega}.
\end{equation}

By choosing $J_{2}=a-h_{2}$ we obtain a perfectly absorbing RSAM,
i.e. $r_{u}=0$ at the frequency $\omega=\delta+\alpha\left(a-h_{2}\right)$.
In between these unsaturated and saturated regimes, the effective
reflectivity can be obtained by solving Eq.~(\ref{eq:RSAM}) numerically.
We represent the results of this procedure for $R=\left|r\right|^{2}$
in Fig.~\ref{RSAM_theo}.

\subsubsection{Defect in the RSAM surface}

We assume that the impurity on the RSAM surface has the effect of
locally lowering the reflectivity of the top mirror. As such the parameters
$a$ and $h_{2}$ are allowed to vary in the transverse dimension.
A decrease of $r_{1}^{\left(2\right)}$of $0.1$ modifies $a$ and
$h_{2}$ to be $a=12$ and $h_{2}=24$. We assume a Gaussian profile
for such transverse variations whose FWHM is $\sim5\,\mu$m.

\subsection{Mode-Locking dynamics}

\begin{figure}[t]
\centering{}\includegraphics[bb=0bp 0bp 300bp 396bp,clip,width=0.45\textwidth]{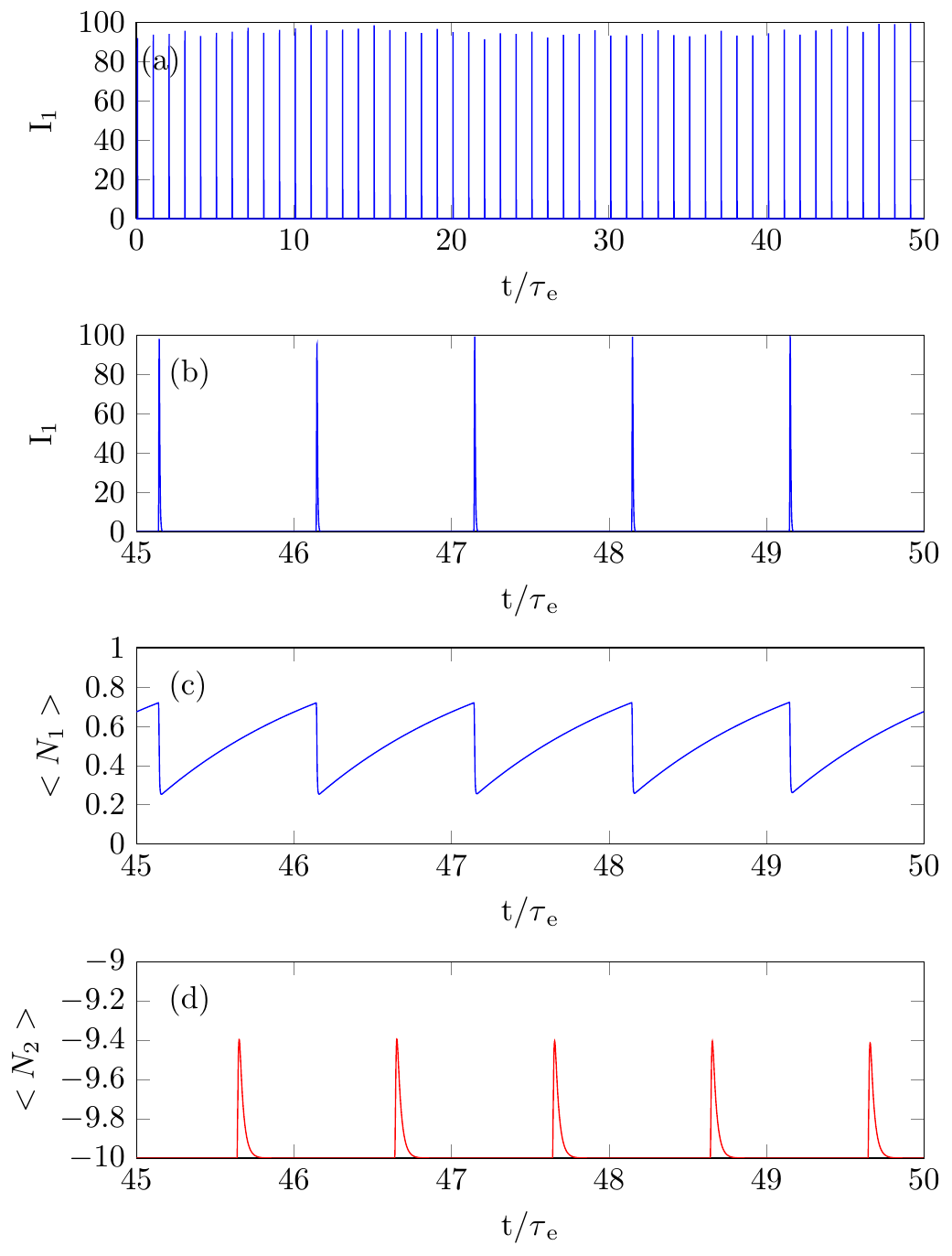}
\caption{Panels a) and b): Averaged Intensity profile $I_{1}$ as a function
of time showing stable PML. Panels c) and d) represent the averaged
population inversion in the VCSEL and in the RSAM as a function of
time. \label{PML_theo-1} }
\end{figure}

The bias current is fixed to $J_{1}=0.92$, i.e. below the threshold
of the solitary VCSEL but above the threshold of the compound device.
We depict in Fig.~\ref{PML_theo-1}, the output time trace for the
intensity averaged over the surface of the VCSEL as well as the averaged
population inversion in the active region of the VCSEL and in the
RSAM. One recognizes the standard PML temporal pattern where the gain
experiences depletion during the passing of the pulse followed by
an exponential recovery. The same yet inverted pattern is also visible
on the RSAM population inversion. We found the pulse width to be of
the order of $4\,$ps, in good qualitative agreement with the experimental
results.

\begin{figure}[t]
\centering{}\includegraphics[bb=0bp 20bp 300bp 460bp,clip,width=0.45\textwidth]{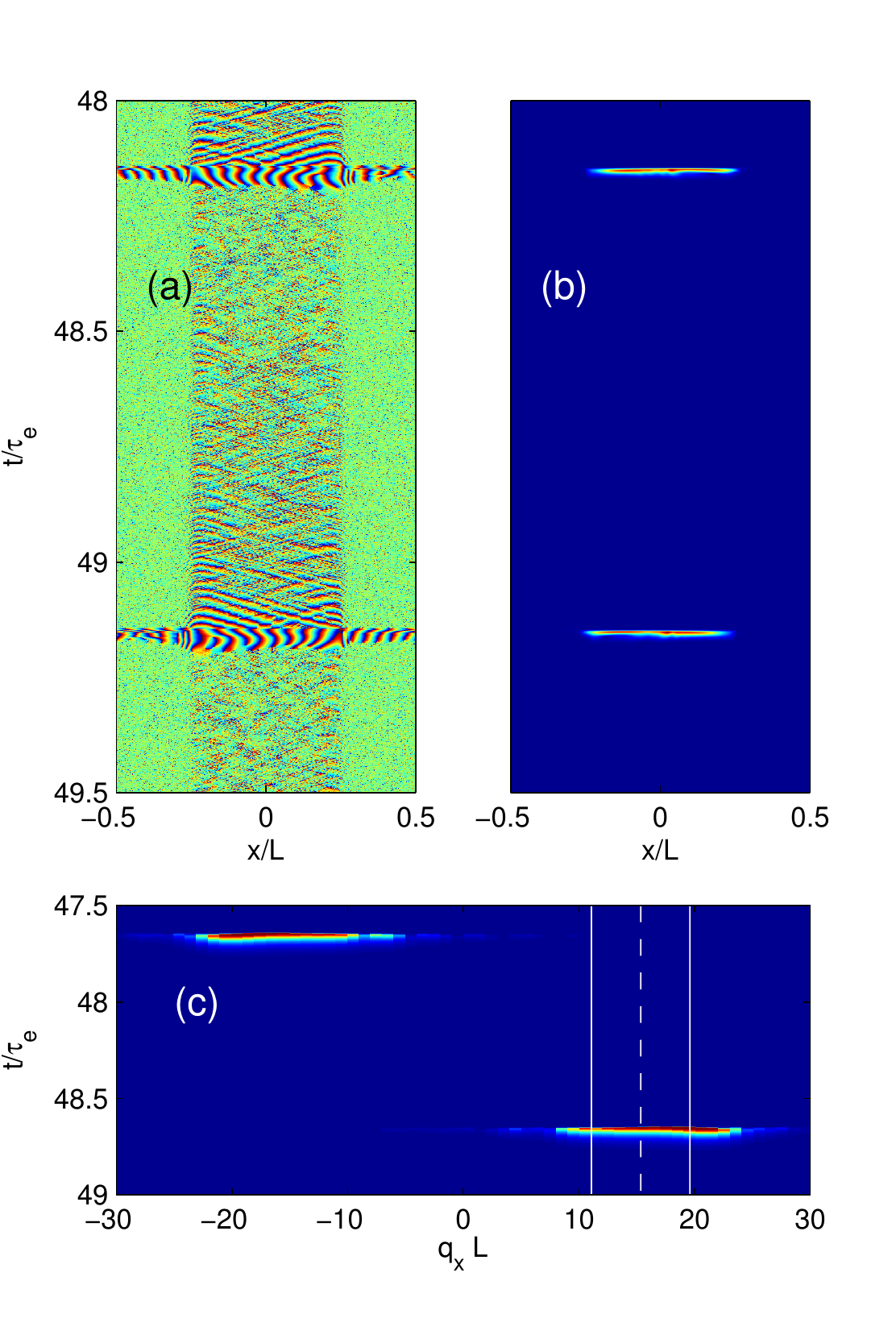}
\caption{Panel a): Phase profile of $E_{1}$ as a function of space and time.
A tilted wave with alternate wave vector from one round-trip to the
next is clearly visible. Panel b) Intensity profile of $E_{1}$ showing
that the transverse intensity profile is uniform. Panel c): Population
inversion in the RSAM. The FWHM size of the population inversion spot
is $\sim6\,\mu$m and the RSAM is fully saturated locally by the pulse.
\label{PML_theo-2} }
\end{figure}

Such averaged representation describes well the experimental situation
when the full output of the VCSEL is captured by the photo-detector.
Yet it does not shed light onto the hidden transverse dynamics. We
represent in Fig.~\ref{PML_theo-2} the spatially resolved temporal
output of the VCSEL. In Fig.~\ref{PML_theo-2}a) the phase profile
$\phi_{1}=\arg\left(E_{1}\right)$ discloses the existence of alternate
transverse waves while the intensity profile in Fig.~\ref{PML_theo-2}b)
remains uniform. Such field profile directly impings the RSAM via
its Fourier Transform which explains that the temporal time trace
of population inversion in the RSAM (Fig.~\ref{PML_theo-2}c)) exhibits
two spots at two opposite locations with respect to its center. In
Fig.~\ref{PML_theo-2}c) we represented the center of the pinning
potential with a dotted line while the two white lines represent its
FWHM. The horizontal axis in Fig.~\ref{PML_theo-2}c) is scaled in
normalized spatial frequencies. The transverse value of the wave vector
here is $\sim16$ which corresponds to $8$ transverse oscillations
along the active region of the VCSEL, i.e. a wavelength of $\sim25\,\mu$m. 

\begin{figure}[t]
\centering{}\includegraphics[bb=30bp 20bp 410bp 320bp,clip,width=0.5\textwidth]{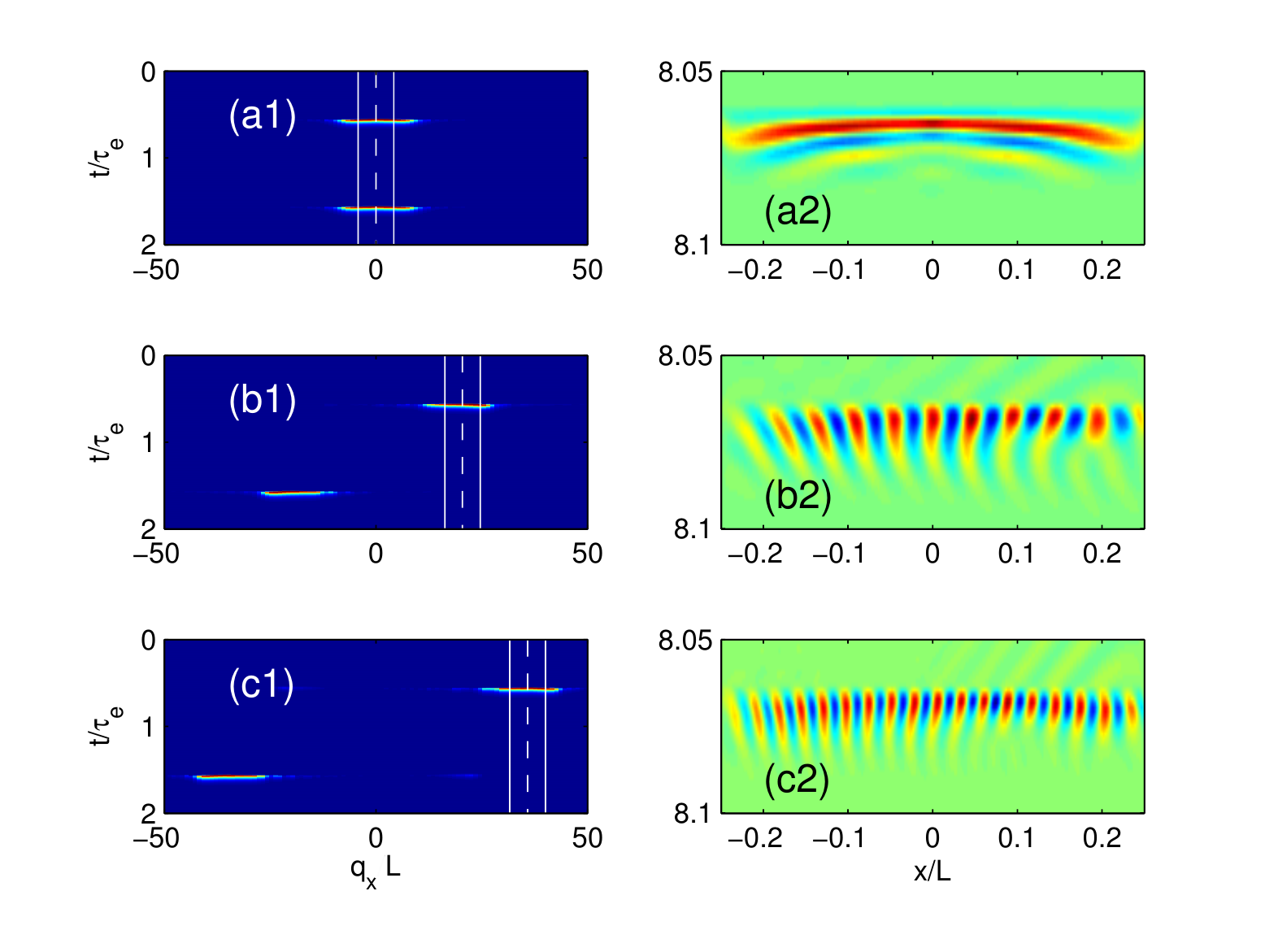}
\caption{Evolution of the transverse wave vector as a function of the position
of the pinning potential. Panels a1), b1) and c1) correspond to the
population inversion in the RSAM while the panels a2), b2) and c2)
correspond to the real part of $E_{1}$. The FWHM size of the population
inversion spot in the RSAM corresponds to $\sim6\,\mu$m and the RSAM
is fully saturated locally by the pulse.\label{PML_theo-3} }
\end{figure}

We found that by slowly displacing the center of the pinning potential
it was possible to tune the normalized transverse wave vector between
$q_{x}L=0$ and $q_{x}L=36$, which corresponds to a minimal wavelength
of $11\,\mu$m. The associated shift in the emission wavelength is
$\sim4\,$nm around $980\,$nm, in excellent agreement with the experimental
results. Several points along such tuning curve are presented in Fig.~\ref{PML_theo-3}.
All cases correspond to stable PML regimes that would seem to be almost
identical if one would considers only the averaged temporal output
like e.g. in Fig.~\ref{PML_theo-1}. Noteworthy, we also found some
perfectly regular PML regimes when the inhomogeneity was located at
the center of the RSAM. We provide several explanations for such discrepancy.
First, the dissipation of the energy and of the associated heat incurred
by the light absorption is doubled when the two spots are well separated
onto the RSAM. Such thermal effects are not taken in consideration
by our model. Second, other spatial inhomogeneities, these ones detrimental
to PML, could very well be located on the sides of the pinning defect
favoring PML. When the two spots become less and less separated experimentally,
the spot that is not experiencing the pinning defect (i.e. the left
spot in Fig.~\ref{PML_theo-3}) will eventually be the victim of
such other detrimental inhomogeneities.

\section{Conclusions}

We have shown that electrically biased broad-area VCSELs with optical
feedback from a RSAM can be passively mode-locked when the VCSEL and
RSAM are placed each at the Fourier plane of the other. In this configuration,
the system emits a train of pulses of $\sim10\,$ps width with a period
equal to the round-trip time of the external cavity $\tau_{e}$. In
this PML regime the time-averaged VCSEL far-field, which is imaged
onto RSAM plane, exhibits two bright peaks symmetrically located around
the optical axis, thus indicating that the VCSEL emits two tilted
waves with opposite transverse components. The time traces corresponding
to each spot consist of a pulse train with a period $2\tau_{e}$,
and the two trains are delayed by one round-trip one with respect
to the other. Accordingly, the two tilted waves are alternatively
emitted at every round-trip. We have shown that the mechanism leading
to wave vector selection is related to the existence of inhomogeneities
in the transverse section of the RSAM. Because the RSAM is in the
Fourier plane of the VCSEL near-field, a defect may favor a tilted
wave emission with a well-defined transverse component. By shifting
the RSAM laterally, the defect moves in the Fourier plane, thus selecting
another transverse wave vector and allowing for wavelength tuning
of the mode-locked emission.

\section{Appendix}

Our approach extends the methodology developed in \cite{MB-JQE-05}
for single-mode VCSELs under optical injection to the case of broad-area
devices. The Quantum-Well active region is considered as infinitely
thin, and the wave equation is exactly solved inside the cavity for
monochromatic plane-waves. Transforming back to a spatio-temporal
representation provides the time-domain evolution of the field inside
cavity. Finally, the coupling of the two cavities is included by describing
how the injection terms depend on the intra-cavity fields.

The starting point is the scalar Maxwell equation for a monochromatic
field as in \cite{MB-JQE-05} 
\begin{eqnarray}
\left(\partial_{z}^{2}+\Delta_{\perp}\right)\mathcal{E}\left(\omega,\vec{r}\right)+\frac{\omega^{2}}{\upsilon^{2}}\mathcal{E}\left(\omega,\vec{r}\right) & \negthickspace\negthickspace=\negthickspace\negthickspace & \frac{-\omega^{2}}{c^{2}\varepsilon_{0}}\mathcal{P}\left(\omega,\vec{r}\right)\,,
\end{eqnarray}
 where $\Delta_{\perp}=\partial_{x}^{2}+\partial_{y}^{2}$, $\mathcal{P}$
is the polarization of the QW active region and $\upsilon=c/n$ with
$n$ the index of refraction. In the longitudinal direction, the cavity
is defined by two Bragg mirrors at $z=0$ and $z=L$.

Fourier transforming over the transverse coordinates yields 
\begin{eqnarray}
\left(\partial_{z}^{2}+\frac{\omega^{2}}{\upsilon^{2}}-q_{\perp}^{2}\right)\mathcal{E}\left(\omega,q_{\perp},z\right) & \negthickspace\negthickspace\negthickspace=\negthickspace\negthickspace\negthickspace & \frac{-\omega^{2}W}{c^{2}\varepsilon_{0}}P\left(\omega,q_{\perp}\right)\delta_{l}.\label{eq:MAX}
\end{eqnarray}
We assumed that the Quantum Well(s) of width $W\ll\lambda$ are located
at $z=l$ and defined $\delta_{l}=\delta\left(z-l\right)$. In the
empty regions where there is no polarization the solution of Eq.~(\ref{eq:MAX})
reads 
\begin{equation}
\mathcal{E}\left(\omega,q_{\perp},z\right)=\begin{cases}
L_{+}e^{iQz}+L_{-}e^{-iQz} & \mbox{if }0<z<l\\
R_{+}e^{iQz}+R_{-}e^{-iQz} & \mbox{if }l<z<L
\end{cases}
\end{equation}
where the longitudinal wave vector $Q\left(q_{\perp},\omega\right)=\sqrt{\frac{\omega^{2}}{\upsilon^{2}}-q_{\perp}^{2}}$
. The boundary conditions at the mirrors and at the QW impose that
\begin{eqnarray}
r_{1}L_{-}+t_{1}^{'}Y & = & L_{+}\\
r_{2}R_{+}e^{iQL} & = & R_{-}e^{-iQL}\\
L\left(l\right) & = & R\left(l\right)=\mathcal{E}\left(\omega,q_{\perp},l\right)\\
\partial_{z}R\left(l\right)-\partial_{z}L\left(l\right) & = & -\frac{\omega^{2}}{\varepsilon_{0}c^{2}}P
\end{eqnarray}
where the primed indexes are for transmission and reflection processes
starting outside of the cavity, $r_{1}$ and $r_{2}$ the top (emitting)
and bottom reflectivities and $Y$ is the amplitude of the external
field impinging on the device. After some algebra the equation linking
$\mathcal{E}$, $P$ and $Y$ is found to be 
\begin{eqnarray}
F_{1}\left(Q\right)\mathcal{E} & = & -\frac{\omega^{2}}{2iQ\varepsilon_{0}c^{2}}\Gamma WP+F_{2}\left(Q\right)Y\label{eq:EY}
\end{eqnarray}
with
\begin{equation}
F_{1}=\left(1-r_{1}r_{2}e^{2iQL}\right),\, F_{2}=t'_{1}e^{iQl}\left(1+r_{2}e^{2iQ\left(L-l\right)}\right).
\end{equation}

The modes of the VCSEL correspond to the minima of $F_{1}$, and the
QW is placed in order to maximize the optical confinement factor $\Gamma=\left(1+r_{1}e^{2iQl}\right)\left(1+r_{2}e^{2iQ\left(L-l\right)}\right)$
for the fundamental mode at $q_{\perp}=0$. For fixed magnitude of
the reflectivities, this is achieved by imposing that
\[
2\frac{\omega}{\upsilon}l+\phi_{1}=2\pi n_{1},\,2\frac{\omega}{\upsilon}(L-l)+\phi_{2}=2\pi n_{2}
\]

\noindent where $\phi_{1,2}$ are the phases of the reflectivities
$r_{1,2}$ and $n_{1}$and $n_{2}$ two integers. Thus the modal frequencies
are determined by the effective cavity length $L_{e}=L+2\upsilon\left(\phi_{1}+\phi_{2}\right)/\omega$.
Around any modal frequency $\omega_{0}$, and in paraxial conditions,
$F_{2}$ and $\Gamma/Q$ vary much more slowly than $F_{1}$. In order
to have a spatio-temporal description of the dynamics of the field,
we then fix $\Gamma/Q$ and $F_{2}$ and expand
\begin{equation}
F_{1}=F_{1}^{0}+(\omega-\omega_{0})\partial_{\omega}F_{1}+q_{\perp}^{2}\partial_{q_{\perp}^{2}}F_{1}+\cdots
\end{equation}

\noindent and transform back to space and time using that $\omega\rightarrow\omega_{0}+i\partial_{t}$
and $q_{\perp}^{2}\rightarrow-\Delta_{\perp}$. Since $F_{1}$ is
at a minimum of its modulus, $\partial_{\omega}F_{1}$ is purely imaginary,
and the field evolution can be written as
\begin{eqnarray}
\tau_{c}\frac{d\mathcal{E}}{dt} & = & i\frac{\omega_{0}\Gamma W}{2n\varepsilon_{0}c}P-\kappa\tau_{c}\mathcal{E}+iL_{diff}^{2}\Delta_{\perp}\mathcal{E}\nonumber \\
 & + & t_{1}^{'}(1+\vert r_{2}\vert)\left(-1\right)^{m}Y+C\Delta_{\perp}^{2}\mathcal{E}
\end{eqnarray}

\noindent where $\tau_{c}\approx\vert r_{1}r_{2}\vert2L_{e}/v$ is
the effective cavity transit time, $\kappa\approx1-\vert r_{1}r_{2}\vert$
are the cavity losses under normal incidence at resonance, $L_{diff}\approx\vert r_{1}r_{2}\vert\frac{2\upsilon\partial_{q_{\perp}^{2}}L_{e}}{\omega_{0}}$
is the diffraction length, and $C$ describes the variation of cavity
losses with the angle of incidence. 

The polarization of the QW active region determines the gain and index
change induced by the carriers, and we adopt the simple adiabatic
approximation
\begin{equation}
P=\varepsilon_{0}(\alpha-i)g_{0}(N-N_{t})E
\end{equation}
where $\alpha$ is Henry's linewidth enhancement factor, $g_{0}$
is the material gain coefficient and $N_{t}$ is the transparency
carrier density. The carrier density, in turn, obeys the standard
description as in \cite{MB-JQE-05}. Upon normalization, the evolution
for the field and carrier density can be written as
\begin{eqnarray}
\frac{\partial E}{\partial t} & = & \left[\left(1-i\alpha\right)N-1+i\Delta_{\perp}+c\Delta_{\perp}^{2}\right]E+hY,\label{eq:E1-1}\\
\frac{\partial N}{\partial t} & = & \gamma\left[J-\left(1+\vert E\vert^{2}\right)N\right]+\mathcal{D}\Delta_{\perp}N,\label{eq:D1-1}
\end{eqnarray}

\noindent where $\gamma$ is the scaled carrier lifetime, $J$ is
the current injection above transparency, $\mathcal{D}$ is the diffusion
coefficient and $h$ describes the coupling of the output field onto
the QW and reads 
\begin{eqnarray}
h & = & t_{1}^{'}\frac{1+r_{2}}{1-r_{1}r_{2}}
\end{eqnarray}

As a last step, we evaluate the field at the laser output $O$ as
a combination of the reflection of the injected beam $r_{1}^{'}Y$
as well as transmission of the left intra-cavity propagating field
$L_{-}$, i.e. $O=t_{1}L_{-}+r_{1}^{'}Y$. Around resonance and using
the Stokes relations $tt'-rr'=1$ and $r'=-r$ we find, defining $\eta=t_{1}/\left(1+r_{1}\right)$
\begin{eqnarray}
O & = & \eta\left(-1\right)^{m}\mathcal{E}-Y
\end{eqnarray}
although the $\left(-1\right)^{m}$ is irrelevant in the sense that
the QW experiences also a field with a $\left(-1\right)^{m}$ which
can therefore be removed. In the case of two devices, we scale the
photon lifetime, the coupling and the saturation field with respect
to the first one leading to the following definitions
\begin{equation}
h_{2}=t_{3}^{'}\frac{1+r_{4}}{1-r_{1}r_{2}}\frac{r_{1}r_{2}}{r_{3}r_{4}}\:,\: a=\frac{1-r_{3}r_{4}}{1-r_{1}r_{2}}\frac{r_{1}r_{2}}{r_{3}r_{4}}\:,\: b=\frac{r_{1}r_{2}}{r_{3}r_{4}}
\end{equation}

and $s=\left(g_{2}\gamma_{1}\right)/\left(g_{1}\gamma_{2}\right)$.

\section*{Acknowledgment}

J.J. acknowledges financial support from the Ramon y Cajal fellowship
and the CNRS for supporting a visit at the INLN where part of his
work was developed. J.J. and S.B. acknowledge financial support from
project RANGER (TEC2012-38864-C03-01) and from the Direcció General
de Recerca, Desenvolupament Tecnològic i Innovació de la Conselleria
d\textquoteright{}Innovació, Interior i Justícia del Govern de les
Illes Balears co-funded by the European Union FEDER funds. M.M. and
M.G. acknowledge funding of Région PACA with the Projet Volet Général
2011 GEDEPULSE.

\bibliographystyle{IEEEtran}
\bibliography{../../BIBLIO/full}

\begin{IEEEbiography}[{\includegraphics[width=2.5cm]{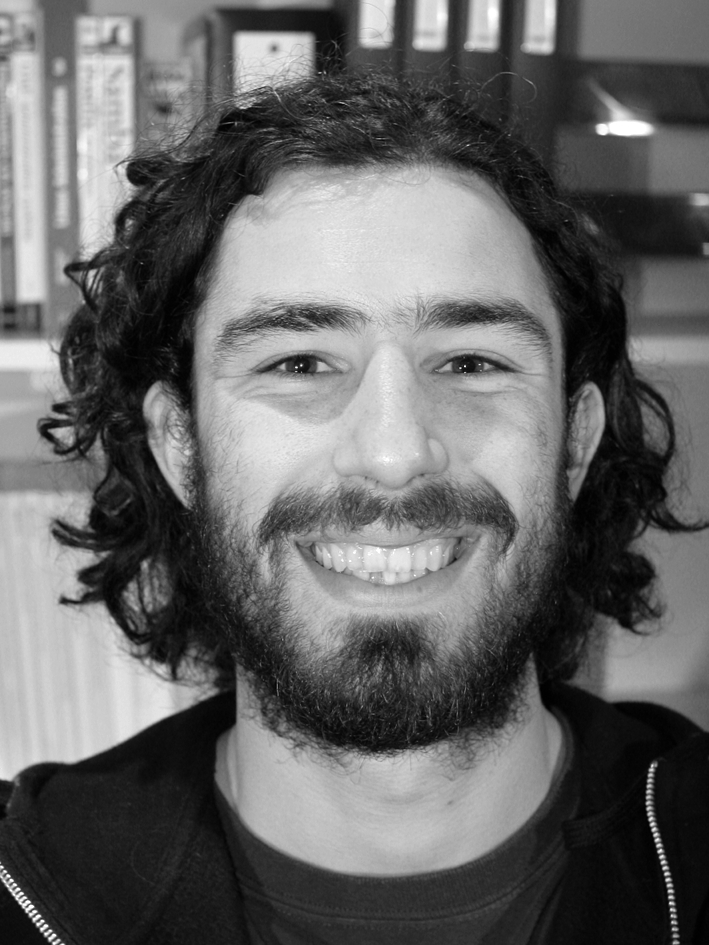}}]
{Mathias Marconi} was born in Nice, France, in 1988. From 2006 to
2011, he was a student at Université de Nice Sophia Antipolis (UNS),
France. In 2011, he was an exchange student at Strathclyde university,
Glasgow, U.K. The same year he obtained the Master degree in optics
from UNS. He is currently pursuing the Ph.D. degree at the Institut
Non-linéaire de Nice, Valbonne, France. His research interests include
semiconductor laser dynamics and pattern formation in out of equilibrium
systems. He is a member of the European Physical Society. 
\end{IEEEbiography}

\begin{IEEEbiography}[{\includegraphics[width=2.5cm]{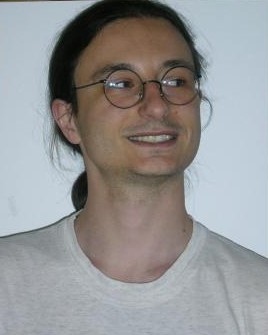}}]
{Julien Javaloyes} (M\textquoteright{}11) was born in Antibes, France
in 1977. He obtained his M.Sc. in Physics at the ENS Lyon the PhD
in Physics at the Institut Non Linéaire de Nice / Université de Nice
Sophia-Antipolis working on recoil induced instabilities and self-organization
processes in cold atoms. He worked on delay induced dynamics in coupled
semiconductor lasers, VCSEL polarization dynamics and monolithic mode-locked
semiconductor lasers. He joined in 2010 the Physics Department of
the Universitat de les Illes Balears as a Ramón y Cajal fellow. His
research interests include laser dynamics and bifurcation analysis.
\end{IEEEbiography}

\begin{IEEEbiography}[{\includegraphics[width=2.5cm]{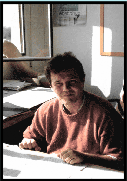}}]
{Salvador Balle}  (M\textquoteright{}92) was born in Manacor, Mallorca.
He graduated in Physics at the Universitat Autònoma de Barcelona,
where he obtained a PhD in Physics on the electronic structure of
strongly correlated Fermi liquids. After postdoctoral stages in Palma
de Mallorca and Philadelphia where he became interested in stochastic
processes and Laser dynamics, he joined in 1994 the Physics Department
of the Universitat de les Illes Balears, where he is Professor of
Optics since 2006. His research interests include laser dynamics,
semiconductor optical response modeling, multiple phase fluid dynamics
and laser ablation. 
\end{IEEEbiography}

\begin{IEEEbiography}[{\includegraphics[width=2.5cm]{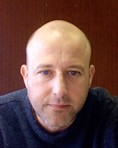}}]
{Massimo Giudici }  (M\textquoteright{}09) received the \textquotedbl{}Laurea
in Fisica\textquotedbl{} from University of Milan in 1995 and Ph.D
from Université de Nice Sophia-Antipolis in 1999. He is now full professor
at Université de Nice Sophia-Antipolis and deputy director of the
laboratory \textquotedblleft{}Institut Non Linéaire de Nice\textquotedblright{},
where he carries out his research activity. Prof. Giudici's research
interests revolve around the spatio-temporal dynamics of semiconductor
lasers. In particular, he is actively working in the field of dissipative
solitons in these lasers. His most important contributions concerned
Cavity Solitons in VCSELs, longitudinal modes dynamics, excitability
and stochastic resonances in semiconductor lasers and the analysis
of lasers with optical feedback. He is authors of more than 50 papers
and he is Associated Editor of IEEE Photonics Journal.  \end{IEEEbiography}

\end{document}